\documentclass[%
 reprint,
 amsmath,amssymb,
 aps,
]{revtex4-2}

\usepackage{graphicx}
\usepackage{dcolumn}
\usepackage{bm}

\begin{document}

\preprint{APS/123-QED}

\title{Rack-integrated quantum dot-based source of single and entangled photons at telecom C-band} 
\author{Michal~Vyvlecka}
 \email{michal.vyvlecka@ihfg.uni-stuttgart.de}
 \affiliation{Institut f\"ur Halbleiteroptik und Funktionelle Grenzfl\"achen, Center for Integrated Quantum Science and Technology (IQ\textsuperscript{ST}) and SCoPE, University of Stuttgart, Allmandring 3, 70569 Stuttgart, Germany}

\author{Raphael~Joos}
 \affiliation{Institut f\"ur Halbleiteroptik und Funktionelle Grenzfl\"achen, Center for Integrated Quantum Science and Technology (IQ\textsuperscript{ST}) and SCoPE, University of Stuttgart, Allmandring 3, 70569 Stuttgart, Germany}
\author{Benjamin~Breiholz}
 \affiliation{Institut f\"ur Halbleiteroptik und Funktionelle Grenzfl\"achen, Center for Integrated Quantum Science and Technology (IQ\textsuperscript{ST}) and SCoPE, University of Stuttgart, Allmandring 3, 70569 Stuttgart, Germany}
\author{Emma~Marmasse}
 \affiliation{Institut f\"ur Halbleiteroptik und Funktionelle Grenzfl\"achen, Center for Integrated Quantum Science and Technology (IQ\textsuperscript{ST}) and SCoPE, University of Stuttgart, Allmandring 3, 70569 Stuttgart, Germany}
\author{Anna~Friederike~K\"ohler}
 \affiliation{Institut f\"ur Halbleiteroptik und Funktionelle Grenzfl\"achen, Center for Integrated Quantum Science and Technology (IQ\textsuperscript{ST}) and SCoPE, University of Stuttgart, Allmandring 3, 70569 Stuttgart, Germany}
\author{Ponraj Vijayan}
 \affiliation{Institut f\"ur Halbleiteroptik und Funktionelle Grenzfl\"achen, Center for Integrated Quantum Science and Technology (IQ\textsuperscript{ST}) and SCoPE, University of Stuttgart, Allmandring 3, 70569 Stuttgart, Germany}
\author{Tobias~Huber-Loyola}
 \altaffiliation[also at ]{Institute of Photonics and Quantum Electronics (IPQ)  and Center for Integrated Quantum Science and Technology (IQST), Karlsruhe Institute of Technology (KIT), Engesserstr. 5, 76131 Karlsruhe, Germany}
 \affiliation{Julius-Maximilians-Universit\"at W\"urzburg, Physikalisches Institut and W\"urzburg-Dresden Cluster of Excellence, ctd.qmat, Am Hubland, W\"urzburg, Bavaria, Germany}

\author{Sven~H\"ofling}
 \affiliation{Julius-Maximilians-Universit\"at W\"urzburg, Physikalisches Institut and W\"urzburg-Dresden Cluster of Excellence, ctd.qmat, Am Hubland, W\"urzburg, Bavaria, Germany}
\author{Michael Jetter}
 \affiliation{Institut f\"ur Halbleiteroptik und Funktionelle Grenzfl\"achen, Center for Integrated Quantum Science and Technology (IQ\textsuperscript{ST}) and SCoPE, University of Stuttgart, Allmandring 3, 70569 Stuttgart, Germany}
\author{Simone~Luca~Portalupi}
 \affiliation{Institut f\"ur Halbleiteroptik und Funktionelle Grenzfl\"achen, Center for Integrated Quantum Science and Technology (IQ\textsuperscript{ST}) and SCoPE, University of Stuttgart, Allmandring 3, 70569 Stuttgart, Germany}
\author{Peter~Michler}
 \affiliation{Institut f\"ur Halbleiteroptik und Funktionelle Grenzfl\"achen, Center for Integrated Quantum Science and Technology (IQ\textsuperscript{ST}) and SCoPE, University of Stuttgart, Allmandring 3, 70569 Stuttgart, Germany}

\date{\today}

\begin{abstract}
For quantum light sources in everyday telecommunication networks, quantum science needs to be fully transformed into quantum technology. The first necessary step to move outside a well-controlled lab environment requires the use of quantum light sources operating in the technologically relevant telecom O- and C-band. This can be provided using epitaxial quantum dots as deterministic sources of quantum light. Particularly intriguing is that emitters operating at telecom wavelengths are rapidly catching up with their short wavelength counterparts in terms of performances.
Here, we make a decisive step forward in the development of quantum communication networks: a state-of-the-art source of quantum light, a semiconductor quantum dot (QD), with record coincidence rate for entangled photon emission in the telecom C-band, is operated inside an optimized rack-based setup. This setup includes a tunable pulsed laser for the QD excitation (from quasi- to fully resonant excitation), all optics for the excitation filtering, and QD signal coupling into single-mode fibers. Overall, the setup allows for above $50\%$ transmission for both exciton and biexciton photons. These results show that quantum dots-based telecom light sources can now be transported and integrated into existing fiber infrastructures, an important step to demonstrate the feasibility of the upcoming quantum internet.

\end{abstract}

\maketitle


\section{Introduction}

The World Wide Web is currently the heart of our global communication, having various distant nodes connected via silica fibers and over satellite links. Quantum technologies are expected to affect everyday life, especially secure communication, quantum sensing, and access to remote quantum computers. This expectation has motivated efforts to move quantum devices from laboratory demonstrations toward practical technologies.~\cite{Wehner2018}. This requires an accurate design optimization of each building block, particularly drastically reducing the overall footprint, ensuring remote control of the setups, and enabling high stability in environments where mechanical vibration and temperature instabilities may be larger than in well-controlled quantum optics laboratory.

When employing quantum light for communication schemes, the two main building blocks will be the source itself, and the receiver station. For quantum light generation, epitaxial quantum dots are raising interest as per their state-of-the-art emission properties as well as for the deterministic nature of the generation process~\cite{MichlerPortalupi2024}. Although interesting experiments have been conducted under electrical pumping~\cite{Bockler2008,Sartison2019,Shooter2020}, optical excitation still provides the highest performance of the emitted photons in terms of single-photon source purity and photon coherence~\cite{Ding2025}. This translates in the need for wavelength-tunable, pulsed mode-locked lasers in all laboratory experiments.

The main limitation of the deployed implementation of QDs stems from the necessity of operating them at cryogenic temperatures (around 4K) in a thermally-stable environment. To partially overcome this limitation, QDs were tested at higher operation temperatures over $40\,$K, allowing their integration in a more compact and affordable Stirling cooler-based cryostats~\cite{Schlehahn2018}. However, higher operation temperatures strongly influence the single-photon source purity, which is the key parameter for long-distance quantum cryptography, and it also influences other parameters such as photon indistinguishability~\cite{Carmesin2018,Nawrath2023}, which is crucial for fundamental quantum operations in photonic quantum repeaters. This motivated the realization of smaller rack-based setups, which demonstrated operation of QDs emitting at wavelengths around $780\,$nm~\cite{Zena2026}.

Here we present a compact, transportable, and rack-based setup that constitutes a source of single- and polarisation-entangled photon pairs at the telecom C-band. This is enabled by the combined use of a telecom InAs/GaAs quantum dot, further operated at 4K to ensure the highest optical performances, with a compact setup including optics control and analysis. Indeed, to be useful in deployed scenarios, the source unit comprises all necessary elements for a fully stand-alone operation of the source, i.e. a compact cryostat, a pulsed laser, optical elements for the photon collection and laser filtering, and remotely controlled optomechanics. From the perspective of the user, the telecom photons are directly available at the output of a single-mode fiber (respectively, two distinct single-mode fibers) for the operation as source of single-photons (source of polarisation-entangled photon pairs) in the telecom C-band. In addition, a second rack is built as a receiver unit, containing state-of-the-art superconducting nanowire single-photon detectors (SNSPDs) and all-optical elements for conducting quantum optical measurements: these span from the estimation of the single-photon nature of the emission (Hanbury-Brown and Twiss setup), quantification of the degree of entanglement (via quantum state tomography) to a quantum cryptographic receiving unit. In the following paper, we will highlight in detail the setup configuration, showing operation as source of single photons and entangled photon pairs, and their use in a field-deployed fiber line across the University of Stuttgart-campus Vaihingen, and the city of Stuttgart, reaching almost $36\,$km with $18\,$dB of loss.

\section{Portable quantum-dot single-photon source}

The central equipment in the portable setup is the compact closed-cycle optical cryostat (attoDRY800xs), pumped by an air-cooled helium compressor. This configuration allows the quantum dot single- and entangled-photon source to operate at a temperature of 4 K without requiring special infrastructure, such as access to cooling water. The optical setup for quantum dot excitation and single-photon collection is divided into three floors: the excitation floor, the collection floor, and the cryostat floor. All three are mounted on independent breadboards positioned above the cryostat vacuum chamber and connected to the cryostat via vibration-damping posts. 

Importantly, all active optical elements are remotely controllable, enabling independent operation of the system at a deployed location, therefore, the user does not need to be physically at the experiment location. The entire system is enclosed in a custom-built transportable box with dimensions of $600 \times 1649 \times 1003\,$mm (W × H × D), mounted on four heavy-duty wheels with a radius of $125\,$mm. The transportable box is equipped with all necessary connections for operating the system in a deployed scenario while fully closed. The control electronics, together with the pulsed fiber excitation laser, are housed in a separate box containing two 19" racks with a total height of 10 standard rack units. When the setup transport box and the electronics box are stacked, the total height of the system is $1980\,$mm, making it compatible with transport through all standard facilities.

\begin{figure*}
  \includegraphics[width=\linewidth]{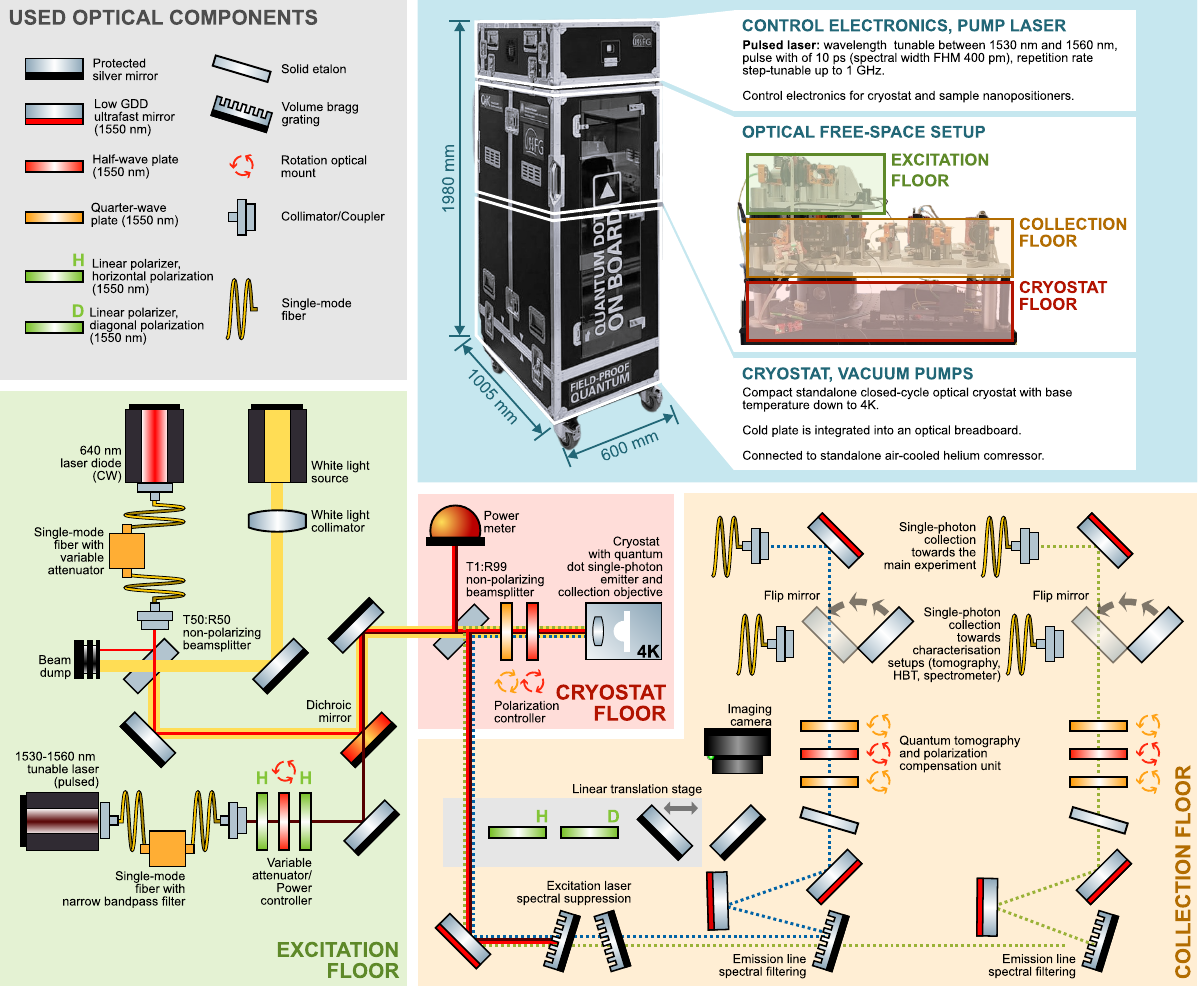}
  \caption{Image and sketch of the transportable source. In the blue block a picture of the overall setup with description of each part. The three floors in the optical setup are color-coded with respect to the sketch below, green of the excitation floor, orange for the collection floor, and red for the cryostat floor. All light sources, optical and optomechanical elements are visualized. In the gray block a legend of the used optical elements is shown.}
  \label{fig:setup}
\end{figure*}

\subsection{Excitation floor}

Key for the operation of a quantum light source based on epitaxial quantum dots is the optical excitation method. In this setup, the quantum dot can be excited by three independent light sources, which are combined using a dichroic mirror into one single spatial mode within the excitation floor (see Fig.~\ref{fig:setup}, green block). i) The first source is a collimated white-light source controlled by an external trigger. It is used for imaging the sample surface, and it provides additional above-band-gap illumination to stabilise the free charge carrier environment within the sample~\cite{Nguyen2012,Michler2017ed}. As shown in the Supplementary Information, the intensity of the two-colour illumination impacts, for the QD used as entangled-photon source, both the blinking and the excitation coherence as measured in the coherent state preparation (similar findings were also observed in droplet-etched GaAs QDs~\cite{Reindl2017}). ii) The second is a compact continuous-wave (CW), fiber-coupled $640\,$nm laser diode which is employed for above-band excitation of the quantum dot. Its output power is controlled using a fiber-coupled MEMS-based variable optical attenuator. The above-band laser beam is combined with the white-light beam into a single spatial mode using a $50:50$ beam splitter.
iii) The third is the primary excitation source, which is a pulsed Er-doped fiber laser with an additional EDFA that provides spectral tunability over the wavelength range from $1530$ to $1560\,$nm; this laser enables the use of various excitation schemes, such as longitudinal acoustic (LA) phonon-assisted excitation or resonant two-photon excitation (TPE). It generates $10\,$ps pulses with a spectral full width at half maximum (FWHM) of $400\,$pm and a repetition rate that can be tuned stepwise from $76\,$MHz up to $1\,$GHz. Spectrally broad background emission originating from the laser fiber amplifier is removed using a fiber-coupled, narrowband optical filter. The laser power is adjusted with a rotatable half-wave plate placed between two aligned polarisers, and it is continuously monitored with a power meter. 

As the combination of the three excitation sources relies exclusively on passive optical components, each source can be controlled independently, while all three can simultaneously illuminate the quantum-dot sample. This capability is crucial for maintaining a stable charge-carrier environment during two-photon excitation measurements.

\subsection{Cryostat floor}

The excitation light from the above floor is coupled into the cryostat chamber (see Fig.~\ref{fig:setup}, red block) through a custom-made non-polarising beam splitter with a splitting ratio of $99:1$. This is used to guide $1\%$ of the excitation to the sample, while the majority of this light is reflected by the beam splitter towards a power meter, which forms part of the active excitation-power stabilisation system. The rear surface of the beam splitter is coated $1050–1700\,$nm (C-coating) to suppress back reflections within the optical setup, while the splitting ratio exhibits extremely low polarisation dependence, making it suitable for the efficient collection of polarisation-entangled photon pairs emitted by the quantum dot. 
The polarisation state of the excitation beam is controlled using motorised quarter-wave and half-wave plates mounted on automatic rotation stages. The beam is then focused onto the sample by a C-coated lens with a focal length of $3\,$mm, placed inside the cryostat chamber. The same lens also serves as the collection optic for the photons emitted by the quantum dot. The $99\%$ reflection ensures the vast majority of the emitted quantum light is sent to the collection floor.

\subsection{Collection floor}

The single photons emitted by the quantum dot are separated from the residual excitation laser by a sequence of spectral filters. The first two polarisation-independent volume Bragg gratings (VBGs) of $1.2\,$nm bandwidth suppress the excitation laser with a suppression ratio of $10^6$ per filter while transmitting the emitted light (see Fig.~\ref{fig:setup}, orange block): this can be done spectrally as in LA-phonon assisted pumping and in TPE, emitted light and excitation laser lay at different wavelengths. The two subsequent polarisation-independent VBGs of $0.55\,$nm bandwidth spectrally separate photons emitted by different quantum-dot transitions, such as the exciton and biexciton transitions, into two distinct spatial modes.

Each photon is further spectrally filtered by a narrow-band angle-tunable optical etalon (Free Spectral Range: $291.6\,$GHz, Finesse: $18.1$), providing additional background suppression and improved spectral selectivity. The photons subsequently pass through a sequence of a quarter-wave plate, a half-wave plate, and a second quarter-wave plate, each mounted on an independently controlled motorised rotation stage. This arrangement enables arbitrary single-qubit polarisation transformations and is used both for quantum state tomography and for polarisation compensation in deployed-fiber experiments (see Fig.~\ref{fig:deployed}). For the polarisation compensation protocol, the polarisation transformation introduced by the deployed optical fiber is characterised using known input polarisation states, namely horizontal and diagonal polarisation. These states are prepared by calibrated linear polarisers mounted on an automated linear translation stage at the beginning of the collection floor.

Following polarisation control, the photons are either coupled directly into single-mode fibers for the desired experiment or redirected by a motorised flip mirror into a characterisation arm. The characterisation setups include a Hanbury-Brown and Twiss setup for second-order correlation measurements, and polarisation-analysis modules for quantum state tomography or the photons can be coupled into an external spectrometer. These characterisation instruments are interconnected using a polarisation-independent MEMS fiber switch, allowing flexible and remote-controllable routing of the collected photons.

The entire collection setup is designed to maximise the single-mode fiber coupling and overall transmission efficiency. This is achieved by maintaining a compact optical layout and employing high-reflectivity and low-group-delay-dispersion (GDD) mirrors throughout the collection path. The low-GDD mirrors are particularly important for preserving the fidelity of polarisation-entangled photon pairs, as they minimise uncontrolled polarisation-dependent phase shifts introduced upon reflection. In addition, the collection floor incorporates a CCD camera for sample imaging and optical alignment. The beam is directed to the camera by a mirror mounted on an automated linear translation stage.

\section{Optical and quantum optical measurements}

In order to demonstrate state-of-the-art operation of the transportable setup as a source of single- and polarisation-entangled photon pairs in the telecom C-band, InAs/GaAs quantum dots have been used as sources of quantum light. Thanks to a metamorphic buffer layer, the strain is engineered to enable emission in the telecom C-band~\cite{Sittig2022}. To increase the source brightness, QDs have been realized within a planar $\lambda$-cavity structure, which consists of two Distributed Bragg Reflector (DBR) stacks: a bottom AlAs/GaAs DBR, and a top dielectric DBR composed of alternating $\text{SiO}_2/\text{TiO}_2$ layers, as in Ref.~\cite{Joos2026} and schematically shown in the inset of Fig.~\ref{fig:sample}~a).

\subsection{Single-photon emission}

\begin{figure*}
  \includegraphics[width=\linewidth]{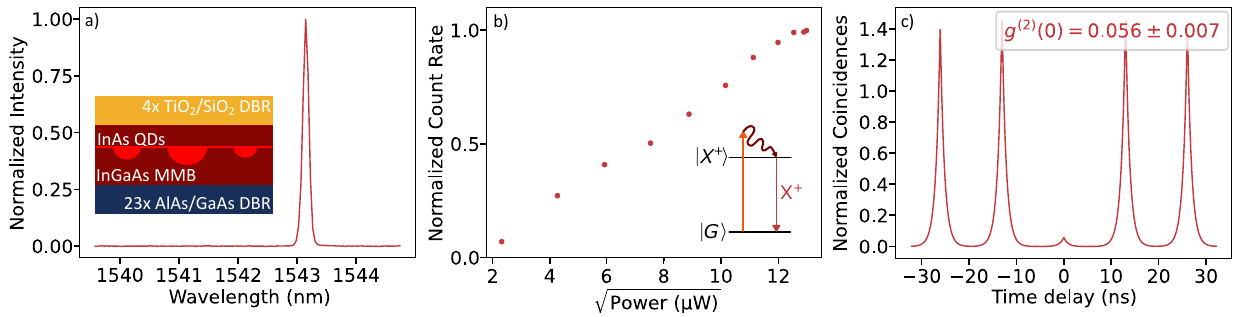}
  \caption{a) Spectrum of one selected quantum dot under LA-phonon assisted excitation. As the laser is fully suppressed by the spectral filtering, only the QD exciton transition is visible. In the inset, a schematic of the planar cavity sample, with a bottom semiconductor DBR ($23$ pairs) and a top oxide DBR ($4$ pairs). InAs QDs are grown on an InGaAs metamorphic buffer layer (MMB) for the strain engineering enabling emission in the telecom bands. b) Intensity over excitation power showing a saturation behavior typical of incoherent excitation methods. In the inset, a schematic energy diagram of LA-phonon assisted process followed by phonon and photon emission. c) Normalized photon autocorrelation measurement, $g^{(2)}(\tau)$.}
  \label{fig:sample}
\end{figure*}

The optical setup was first characterised using a quantum dot operated under LA-phonon-assisted excitation~\cite{LAParis:21}. This excitation scheme enables efficient spectral suppression of the excitation laser while providing high robustness against excitation-laser power fluctuations and a low multi-photon emission probability~\cite{LAParis:21,Reindl2017,Axt:PRL19}. These properties are particularly important for the implementation of quantum-dot single-photon sources in deployed quantum communication and quantum cryptography protocols~\cite{Vyvlecka2023}.

A quantum-dot transition at a wavelength of $1543\,$nm, further referred to as QD-LA, was excited by a laser spectrally blue detuned by $1.5\,$nm from the transition (see inset of Fig.~\ref{fig:sample}~b)), and operated at a repetition rate of $76\,$MHz. The emission spectrum of QD-LA after the full suppression of the excitation laser is shown in Fig.~\ref{fig:sample}~a).

The measured dependence of the single-photon count rate on the excitation power exhibits the characteristic behaviour of LA phonon-assisted excitation, reaching a saturation plateau at higher excitation powers (Fig.~\ref{fig:sample}~b)). The normalized single-photon count rate was extracted from the second-order correlation measurements. At saturation, a count rate of $1.3\,$Mcps was measured in the single-mode fiber-coupled main collection port, corresponding to an overall end-to-end system efficiency of approximately $2\%$.

The single-photon nature of the emission was verified by measuring the second-order correlation function, $g^{(2)}(\tau)$. At saturation, the zero-delay value was measured to be $g^{(2)}(0)=0.056 \pm 0.007$, confirming strong suppression of residual excitation laser (see Fig.~\ref{fig:sample}~c)). This value can be further reduced by optimising the excitation conditions, particularly the spectral detuning and excitation power~\cite{Axt:PRL19,Vyvlecka2023,Jehle2026}. For example, at an excitation power corresponding to $75\%$ of the saturation power, a value of $g^{(2)}(0)=0.031 \pm 0.002$ was obtained, demonstrating an even lower multi-photon probability at the expense of a single-photon count rate reduced to $86\%$ of its saturation value. The QD-LA's emission shows pronounced blinking; see the supplementary information. The QD-LA was active for $60.9\%$ of the time.

\subsection{Entangled-photon pair emission}

\begin{figure*}
  \includegraphics[width=\linewidth]{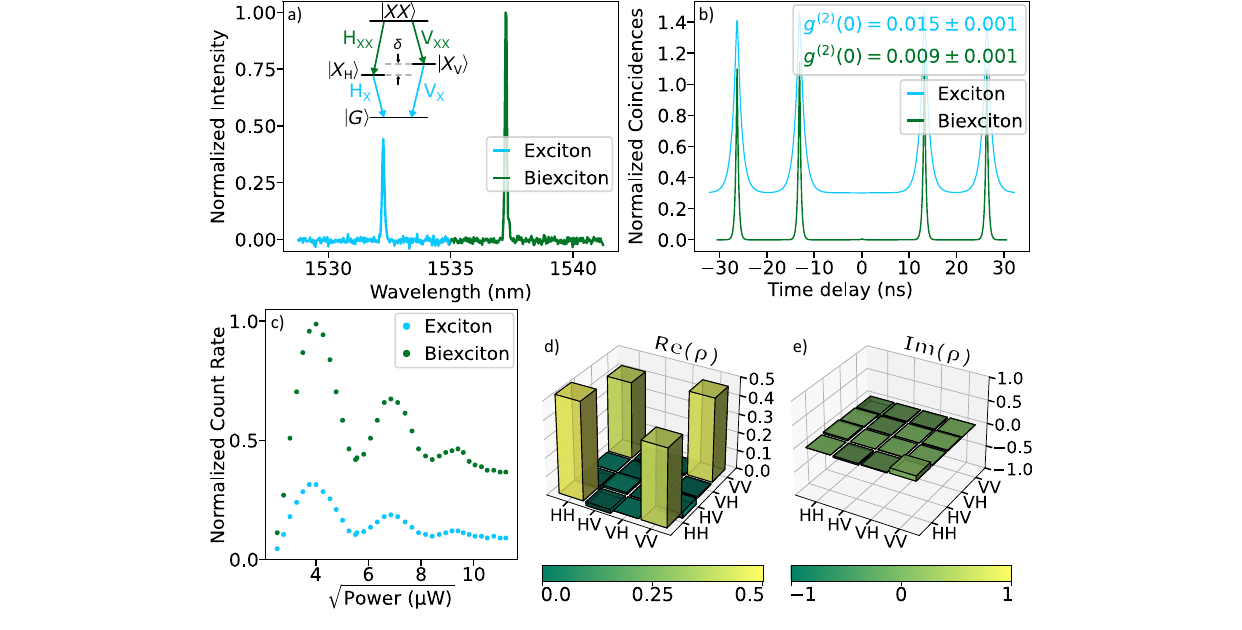}
  \caption{a) Spectrum of one selected quantum dot under TPE excitation. As the laser is fully suppressed by the spectral filtering, only the QD exciton and biexciton transitions are visible. b) Normalized photon autocorrelation measurements, $g^{(2)}(\tau)$, for the exciton and biexciton emission, where the exciton measurement is shifted by an offset of 0.3. c) Intensity over excitation power showing Rabi oscillations typical of coherent excitation methods for the exciton and biexciton emission.d) Real part and e) imaginary part of the reconstructed two-photon density matrix obtained by quantum state tomography for the emitted polarization-entangled photon pairs, using a coincidence time window of $10\,$ps.}
  \label{fig:TPE}
\end{figure*}

A second QD located on a second chip of a similar design (where instead the bottom DBR is semiconductor, while the top oxide-based DBR has lower reflectivity), hereafter referred to as QD-TPE, was investigated under TPE. This excitation scheme enables the generation of polarisation-entangled photon pairs through the biexciton-exciton radiative cascade~\cite{Mueller2014}. Such entangled photon pairs represent a fundamental resource for entanglement-based quantum cryptography protocols as well as for entanglement swapping, which is a key building block of photonic quantum repeaters. Furthermore, resonant TPE leads to emission of photons with an ultra-low multi-photon contribution~\cite{Fisher:NPJ18}, which is essential for secure quantum communication, particularly for long communication distances and in mistrustful quantum cryptographic protocols~\cite{Bozzio2022}.

The exciton and biexciton emission lines are centred at wavelengths of $\lambda_\text{X} = 1532.1\,$nm and $\lambda_\text{XX} =1537.1\,$nm with respective linewidths $\Gamma_\text{FWHM, X}=6.93(12)\,$GHz and $\Gamma_\text{FWHM, XX}=4.27(3)\,$GHz~\cite{Joos2026}. The emission spectrum after suppression of the excitation laser is shown in Fig.~\ref{fig:TPE}~a). During TPE, the free-charge-carrier environment of the sample was stabilised by weak additional white-light illumination. The white-light intensity was optimised to minimise emitter blinking and thereby maximise the detected photon count rate. In addition, the two-colour illumination improved the fidelity of coherent state preparation (see Supplementary Information).

After correcting for efficiencies of detection and the quantum tomography setup, the fiber-coupled single-photon count rates were measured to be $5.04 \pm 0.16\,$Mcps for the biexciton and $1.97 \pm 0.06\,$Mcps for the exciton at an excitation repetition rate of $76\,$MHz as previously reported in Ref.~\cite{Joos2026}. These values correspond to end-to-end system efficiencies of approximately $6.6\%$ and $2.6\%$ for the biexciton and exciton photons, respectively. The QD-TPE is optically active for $81\%$ of the time, see the supplementary information. The single-photon character of both transitions was confirmed by measurements of the second-order correlation function, yielding zero-delay values of $g^{(2)}(0)=0.009 \pm 0.001$ for the biexciton and $g^{(2)}(0)=0.015 \pm 0.001$ for the exciton (Fig.~\ref{fig:TPE}~b)). The measured Rabi oscillations (Fig.~\ref{fig:TPE}~c)) further demonstrate the high fidelity of coherent state preparation.

In addition to the high fiber-coupled coincidence rate of $201 \pm 13\,$kcps~\cite{Joos2026} and the exceptionally low multi-photon emission probabilities, QD-TPE exhibits a fine-structure splitting of only $3.2\,\mu\text{eV}$, providing the essential prerequisite for the generation of high-fidelity polarisation-entangled photon pairs. A maximum entanglement fidelity of $0.964$ was obtained for a $5\,$ps coincidence time window~\cite{Joos2026}. The corresponding reconstructed density matrix for a coincidence window of $10\,$ps is shown in Fig.~\ref{fig:TPE}~d) and e).

\subsection{Deployed quantum communication}

\begin{figure*}
  \includegraphics[width=\linewidth]{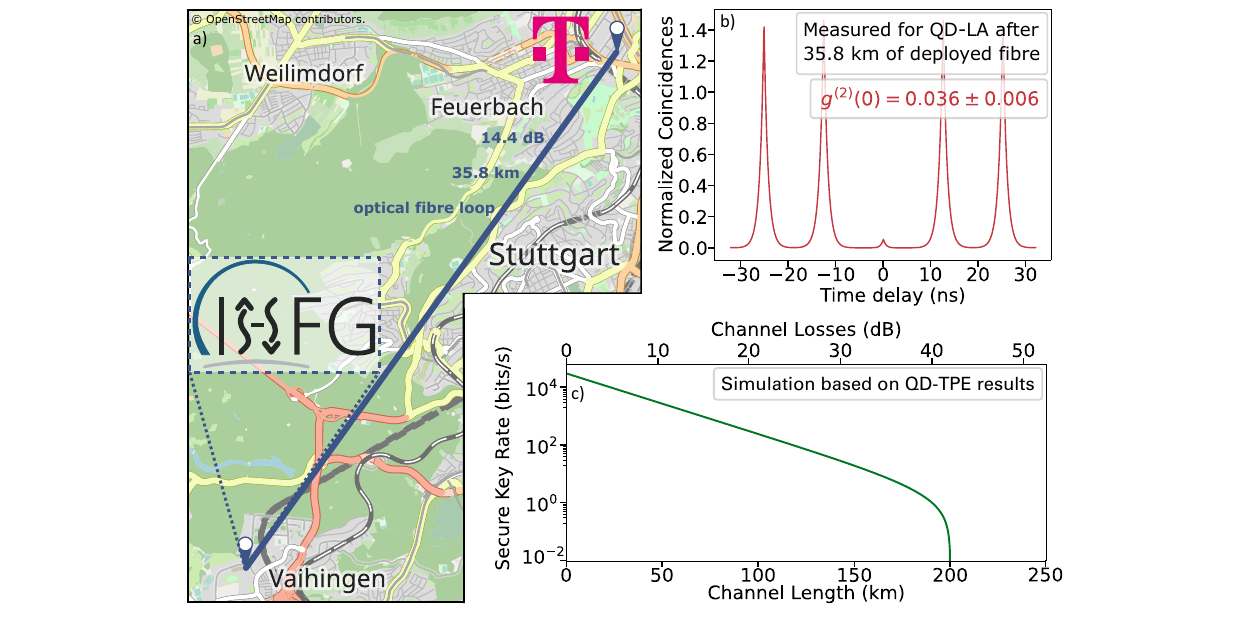}
  \caption{a) Deployed metropolitan optical fiber link in Stuttgart, Germany, connecting the research laboratory of the Institut f\"ur Halbleiteroptik und Funktionelle Grenzflächen (IHFG) on the Vaihingen campus of the University of Stuttgart with the Deutsche Telekom AG site in Stuttgart-Feuerbach. The link consists of two optical fibers connected in Stuttgart-Feuerbach to form a fiber loop. The total loop length is $35.8\,$km with a propagation loss of $14.4\,$dB. The displayed map is based on OpenStreetMap data; for copyright information, see openstreetmap.org/copyright. b) Normalized second-order autocorrelation measurement, $g^{(2)}(\tau)$, of QD-LA after transmission through the deployed metropolitan fiber loop. c) Theoretically estimated secure key rate of the BBM92 quantum key distribution protocol using QD-TPE as a function of communication distance. The calculation assumes a detector efficiency of 0.8 and a fiber attenuation of $0.21\,$dB/km.}
  \label{fig:deployed}
\end{figure*}

The potential of the transportable quantum-dot-based source of single and entangled photons for future deployed quantum communication experiments was first demonstrated by connecting the QD-LA source to a deployed metropolitan single-mode fiber loop. The fiber link connects the laboratory at the Institut für Halbleiteroptik und Funktionelle Grenzflächen (IHFG) on the Vaihingen campus of the University of Stuttgart with a facility of Deutsche Telekom AG in Stuttgart-Feuerbach, where the fiber loop is turned back towards the laboratory (Fig.~\ref{fig:deployed}~a))~\cite{strobel2024}. The total fiber length is $35.8\,$km with an overall transmission loss of $18\,$dB, the majority of which originates from fiber connectors rather than fiber attenuation.

Single photons emitted by QD-LA under LA phonon-assisted excitation were transmitted through the deployed fiber loop. After correcting for the detector efficiency, a single-photon count rate of $20\,$kcps was measured at the output of the link. The single-photon source purity remained unaffected by the transmission, yielding a zero-delay second-order correlation value of $g^{(2)}(0)=0.036 \pm 0.006$, as shown in Fig.~\ref{fig:deployed}~b). These results demonstrate that the source is well suited for deployed quantum communication applications, including quantum key distribution (QKD)~\cite{Pan:RevMod20,Bozzio2022}.

The potential of the QD-TPE source for long-distance entanglement-based QKD, specifically the BBM92 protocol~\cite{Bennett1992}, was further investigated by numerical simulations based on the experimentally measured source characteristics. The secure key rate (SKR) was calculated using the measured time-resolved density matrices as the ones in Fig.~\ref{fig:TPE}~d) and e)) together with the experimentally determined coincidence rates~\cite{Schimpf2021}. For each communication distance, the coincidence time window was optimised to maximise the SKR (Fig.~\ref{fig:deployed}~c)).

As an example, for a communication fiber length of $175\,$km, corresponding to a total channel propagation loss of $36.75\,$dB in single-mode silica optical fiber, the calculated maximum SKR reaches $4.2\,$cps. To the best of our knowledge, this value exceeds all previously reported SKR values demonstrated for deployed entanglement-based quantum QKD using QD entangled-photon sources~\cite{BassoBasset2026}, highlighting the strong potential of the presented source for future long-distance quantum communication networks.

\section{Conclusion}

In this work, we demonstrate a compact, transportable source of single- and entangled-photon pairs operating in the telecom C-band. The setup includes all necessary elements, from the cryostat to excitation lasers and optics, for operating the telecom C-band QD sources at any location without needing dedicated infrastructure. Thanks to the automatic control of all active elements, the user can run the experiment from a remote location. This compact optical setup enables operation of the quantum-dot source without compromising the overall transmission efficiency of the system, which exceeds $65\%$ after correcting for the detector efficiency. This value is indeed higher as compared to highly efficient telecom C-band table-top setups (with previously reported values of circa $30\%$~\cite{Joos2024}). Additionally, the source performances are also not impacted by operating in the rack-based unit. This high setup efficiency together with a bright source of quantum light allowed setting a new record in the generation of polarisation-entangled photon pairs as reported in Ref.~\cite{Joos2026}. Here, this source of entangled photons has been further investigated and showed an exciting potential in its future use for long-distance QKD protocols based on entanglement. From on the observed count rate and estimating a reasonable channel loss, the maximum calculated secret key rate exceeds any previous entanglement-based QKD experiments employing quantum dots. Finally, to prove that such setup can be integrated into existing optical fiber links, a source of single photons was interfaced with an intra-city deployed fiber link of circa $36\,$km without compromising the single-photon nature of the QD emission.
In conclusion, the results show the possibility of realizing a high-efficiency, compact and transportable setup operating with state-of-the-art quantum dots as sources of single- and polarisation-entangled photons in the telecom C-band. This will play a major role in transforming quantum optics experiments into field-deployed quantum technology.

\begin{acknowledgments}
The authors acknowledge Marc~Geitz and Ralf-Peter~Braun from Deutsche Telekom for their support.
M. V., R. J., P. V., M. J., S. L. P. and P. M. acknowledge funding by the German Federal Ministry of Research, Technology and Space (BMFTR) via project QR.X (16KISQ013) and QR.N (16KIS2207). Additional funding was also provided via the project EQSOTIC. This project was funded within the QuantERA II Programme that has received funding from the EU’s H2020 research and innovation programme under the GA No 101017733, and with funding organization BMFTR (with project number 16KIS2060K). T. H.-L. acknowledges financial support from the BMFTR within the Project Qecs (Grant No. FKZ: 13N16272). S.H. acknowledges financial support by the BMFTR through the projects QR.X (FKZ: 16KISQ010), QR.N (FKZ. 16KIS2209) and MHLASQU (FKZ: 13N16027).
\end{acknowledgments}


\bibliography{References}

\end{document}